\documentstyle[aps,prl,balanced,epsfig]{revtex}

\newcommand{\Fig}[1]{Fig.~\ref{#1}}
\newcommand{\Equ}[1]{(\ref{#1})}
\newcommand{\sizeps}{2.65in}

\begin{document}

\title{
Noise induced transition from an absorbing phase to a regime of
stochastic spatiotemporal intermittency
}
\author{M. G. Zimmermann, R. Toral, O. Piro, M. San Miguel}
\address{
Instituto Mediterr\'aneo de Estudios Avanzados (CSIC-UIB)\cite{url}, \\
Campus Universitat Illes Balears, \\ E-07071 Palma de Mallorca,
Spain }

\maketitle
\begin{abstract}    
We introduce a stochastic partial differential equation capable
of reproducing the main features of  spatiotemporal
intermittency (STI).  Additionally the model displays a noise
induced transition from laminarity to the STI regime. We show by
numerical simulations and a mean-field analysis that for high
noise intensities the system globally evolves to a uniform
absorbing phase, while for noise intensities below a critical
value spatiotemporal intermittence dominates. A quantitative
computation of the loci of this transition in the relevant
parameter space is presented.
\end{abstract}    


\begin{twocolumns}

Spatiotemporal chaos (STC) is a complex behavior, common to
many spatially extended nonlinear dynamical 
systems\cite{Hoh89,Cross94,Bohr98,Ahlers98}. This behavior is
characterized by a combination of chaotic time evolution and
spatial incoherence made evident by correlations decaying both
in space and time.
In spite of considerable theoretical and experimental effort
devoted to give a precise definition of STC and its different
regimes, the present status is still unsatisfactory. 
A possible strategy to make progress in the understanding of STC
is  to investigate scenarios based on simple  models, with few
controlled ingredients,  that reproduce the spatio-temporal
structures under study. Such models are instrumental in
searching for generic  mechanisms leading to such complex
behavior.  
Among the available scenarios, few of them consider the
framework  of stochastic partial differential equations
(SPDE)\cite{sancho,Tu97,muno98} to describe STC.  
A successful example is, however, the mapping of the
Kuramoto-Shivashinsky equation (describing a STC regime named
{\it phase turbulence})  to the stochastic model of surface
growth known as Kardar-Parisi-Zhang equation\cite{KPZ}. 

A particular instance of STC is a regime called spatiotemporal
intermittency (STI) which is present in a large variety of
systems\cite{chate,prl97,argen97,zimm97}. Generally speaking,
this regime is a chaotic spatiotemporal  evolution (the {\it
turbulent} phase) irregularly and continuously interrupted by 
the spontaneous formation of domains with a wide range of sizes
and lifetimes, where the behavior is ordered ({\it laminar}).
The borders of the laminar regions propagate as fronts and
eventually cause the collapse of the corresponding region into
the turbulent background. 
There are strong indications that
the STI regime has many features in common with phenomena of
probabilistic nature. 
For example, it appears in some systems that critical exponents
at the onset of STI are in the universality class of directed
percolation\cite{Rolf98}. STI has also been related to
nucleation\cite{argen97}, another process associated with
stochastic fluctuations. However, no description of STI
in terms of SPDEs has been put forward so far.

The purpose of this Letter is to introduce a model, entirely
based on a simple SPDE, that describes the main features of STI,
and report the existence of a noise induced transition from
laminarity to STI.  The laminar phase is associated to an
equilibrium state called {\it absorbing} in the SPDE parlance. 
The role of the turbulent phase is played by a strongly
fluctuating state driven by noise. For either large values of
the noise intensity or small values of the diffusion constant, 
the system is globally attracted to the absorbing
state. However, for a fixed diffusion rate, STI sets in below a
critical noise intensity.  We investigate the nature of this
transition both analytically using a mean-field scheme, and
numerically by means of indicators such as the change of the
mean velocity of fronts between absorbing and turbulent regions,
the one-site probability distribution function and the order
parameter characterizing the average system fraction in the
laminar phase.

We consider the following one-dimensional It\^o-SPDE
\cite{vankamp} for the evolution
of a real field $u(x,t)$ \cite{Itonumerics}:
\begin{equation}
u_t = -\frac{\partial \phi}{\partial u}  + D \, u_{xx} + 
\sqrt{\epsilon}\; g(u) \;\xi
\label{pde}
\end{equation}
The r.h.s. of \Equ{pde} is composed of a gradient term
derived from a  potential $\phi(u)$,  a diffusive
spatial coupling with diffusion constant $D$, and a
multiplicative  Gaussian white noise $\xi(x,t)$ of
zero mean and correlations
$\langle \xi(x,t)\xi(x',t')\rangle\;=2\; \delta(x-x')\delta(t-t')$.  
For the results described below, the essential aspects of our
model are (i) a {\em bistable} potential $\phi(u)$ and (ii) a
multiplicative noise function $g(u)$ which vanishes at the
metastable state.
Different forms of the noise amplitude $g(u)$ 
give rise to different universality
classes\cite{muno98}.

We make the simplest choice for the function
$g(u)=u$ \cite{CommentUniver} so that  the sign of  $u(x,t)$ is
preserved during the evolution of our system, allowing us to
restrict our attention to the case $u(x,t) \ge 0$, by picking
positive defined initial conditions. With this phase-space
restriction, we choose the potential  $\phi(u)= a\, u^2 -u^4 +
h\, u^6$  (with $h>0$), which for  $0<a<a_M=1/4h$ has a relative
minimum at $u=0$ and an absolute minimum at $u=u_+$ separated by
a maximum at $u=u_-$\cite{details}.
At $a=a_M$ we are at the ``Maxwell point'' where $\phi(0)=\phi(u+)$,
while at $a=0$ the $u=0$ state switches from metastable to
unstable.
In the absence of noise a front between a region of $u=0$  and a
region of $u=u_+$ moves at finite speed towards the former, due
to diffusive coupling.    Consequently, any finite region of
$u=0$ surrounded by $u_+$ will eventually shrink letting the
system evolve into  the uniform $u_+$ attractor.
This is the case except for initial configurations laying
completely on the relatively small basin of attraction of 
$u=0$.

\begin{figure}
\centerline{\epsfig{figure=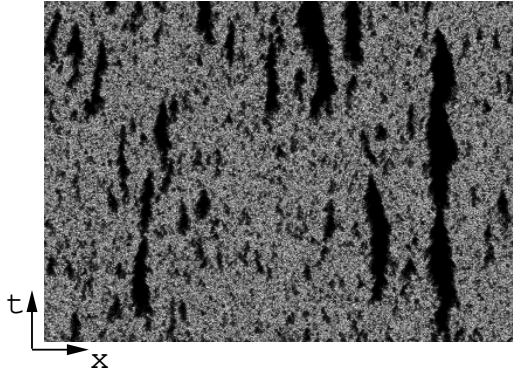,width=\sizeps}}
\caption[]{Space-time evolution of $u(x,t)$  
in a situation where the STI phase persists for all times 
(black: $u=0$, grey: $u>0$). Space ranges in  $x\in(0,400)$
(with periodic boundary conditions) and  time in $t\in(0,90)$.
The numerical method  uses a finite difference approximation for
the Laplacian, and   an   Euler algorithm for the time
integrator which respects the positivity of the field.   The
initial condition is chosen randomly in the interval $u_0(x) \in
(0,2.4)$.  Other values of the parameters are
$\epsilon=0.95$, $a=0.5$, $D=2.0$, $h=0.22$, $\Delta x=1$,
$\Delta t=0.001$.
\label{pdesim}}
\end{figure}

In the presence of multiplicative noise, the  state $u(x)\equiv 0$
becomes an {\em absorbing} state\cite{muno98}, i.e. a state
where the system can be driven to by fluctuations but not
removed from by neither these nor by the deterministic
dynamics. This is because the selected noise amplitude function
$g(u)$ vanishes at the fixed point
$u=0$.  On the other hand, when the field takes values near
$u_+$,  the absolute minimum of $\phi(u)$,  fluctuations are
always active. 
However, without diffusion ($D=0$) the fate of the system is to
asymptotically settle in the absorbing state, since  for any
$\epsilon\neq 0$ a fluctuation large enough to push the system to the
basin of attraction of $u=0$ will eventually   occur.

From this analysis, we infer that noise and
diffusion play opposite roles on the asymptotic space-time
evolution of the field $u$.  While noise nurtures the
development of regions dominated by the absorbing state, 
diffusion favors the 
dominance of $u$ values fluctuating around the global minimum
of $\phi(u)$.  We associate the quiescent uniform absorbing
state of this system with laminarity and the disordered
fluctuations around the global potential minimum with turbulence.
In \Fig{pdesim} we show  a space-time plot of a  numerical
solution of Eq. \Equ{pde}. Notice that for these values of
$\epsilon$ and $D$ there are occasional nucleations
of regions of the laminar or absorbing state that eventually collapse under
the progress of  the fluctuating or turbulent state. The wide
range of sizes and lifetimes of the laminar areas, characteristic
of STI, is evident in the picture.

A similar mechanism, also involving the nucleation of a laminar
state, has been invoked to be responsible of STI in a
deterministic reaction-diffusion system\cite{argen97}. In this
case, the role of noise is played by a chaotic dynamics
generated via a Hopf bifurcation.

\begin{figure}
\centerline{\epsfig{figure=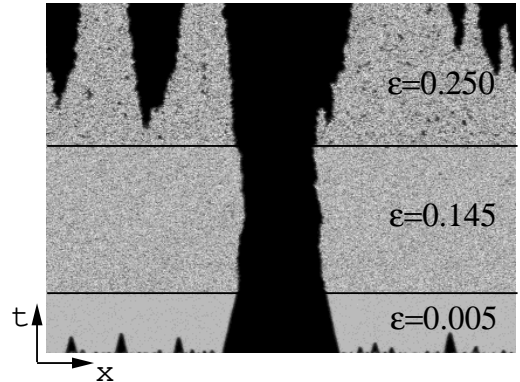,width=\sizeps}}
\caption[]{Space-time plot of $u(x,t)$ starting from a
random initial condition $u_0(x)$ of mean $u_+$ for $x\in(0,400)$ except 
in the interval $x\in(150,250)$ where $u_0(x)=0$. The noise intensity
$\epsilon$ is changed during the evolution:
$\epsilon=0.005$ for $t\in(0,30)$, $\epsilon=0.145$ for
$t\in(30,105)$ and $\epsilon=0.250$ for $t\in(105,180)$. 
Other parameters are $a=1.0$, $D=2.0$. 
\label{front}}
\end{figure}

In \Fig{front}, for fixed $a$ and $D$,  we
compare  the evolution of fronts separating absorbing and active
regions for three  different noise levels.
For the lowest noise intensity $\epsilon$
 these fronts invade the absorbing phase, as in the
deterministic system,
and the system asymptotically tends to be globally in 
the active phase $u_+$.  On the other extreme, for the highest
$\epsilon$, these fronts invade on average  the active phase
which consequently tends to disappear.  The evolution for an
intermediate value of $\epsilon$ shows a situation where the
fronts are at rest on average\cite{Rem_fronts}.
This suggests a possible mechanism for a noise induced transition  
from the STI phase to the absorbing phase.

In \Fig{numchar}a we give a quantitative evidence of this transition.
There the average fraction $R$ of the system in
the absorbing phase is plotted as a function of the noise
intensity $\epsilon$. This figure shows a transition 
from $R \approx 0$ (STI) to $R=1$ (absorbing phase)  at roughly
the same critical noise intensity as for the
front velocity reversal. 
Another useful measure to characterize this transition is the
one-site stationary probability density function
(pdf)\cite{muno98}, $P(u)$. In \Fig{numchar}b we plot $P(u)$ for
different values of the noise intensity. 
We observe a transition from a pdf which has a  hump
around the active phase $u_+$, to a pdf which is highly peaked
at the absorbing state $u=0$. Notice that this transition occurs
at the  value of $\epsilon$ for which $R$ also changes abruptly. 

\begin{figure}
\centerline{\epsfig{figure=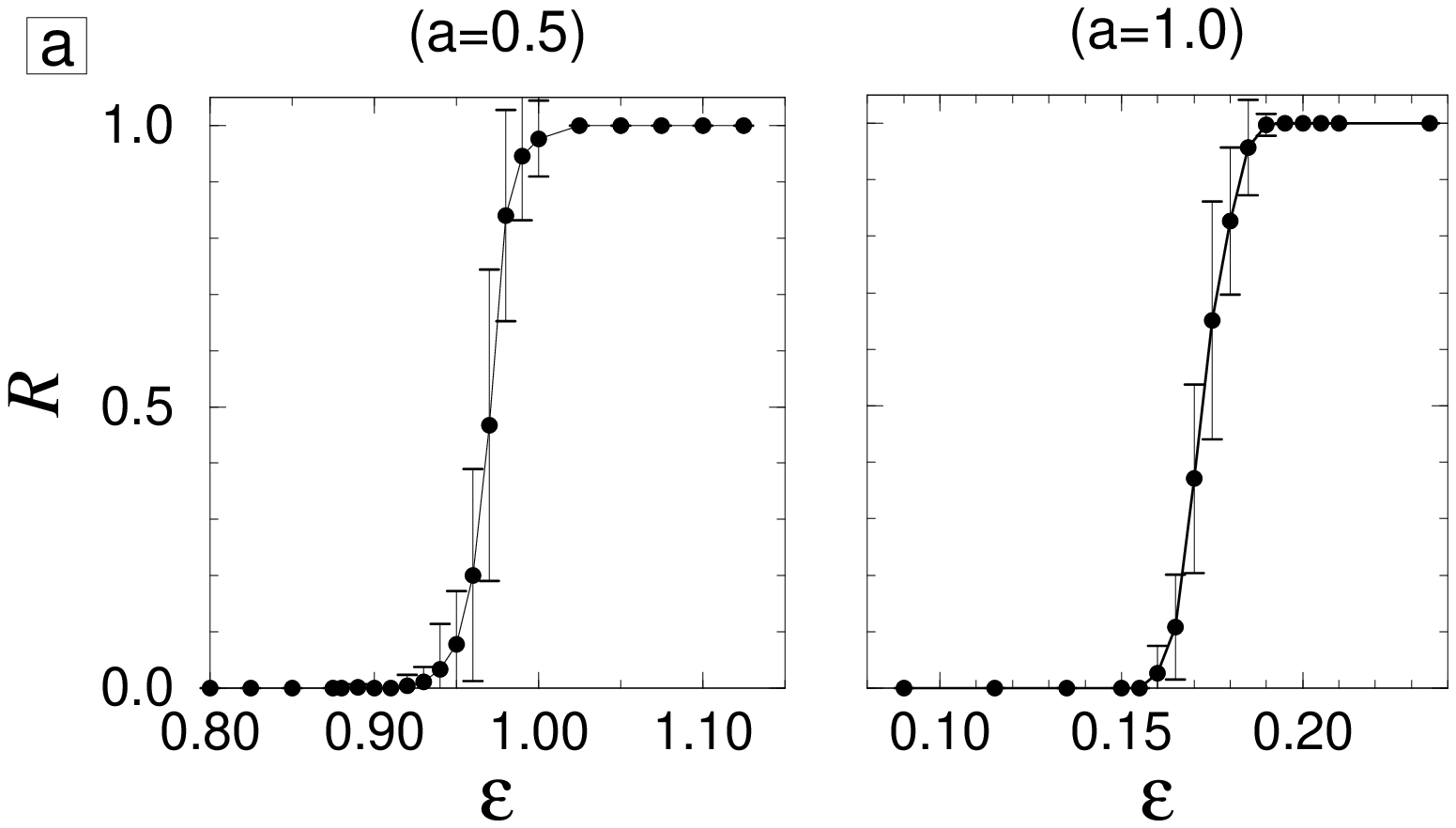,width=2.99in}}
\epsfig{figure=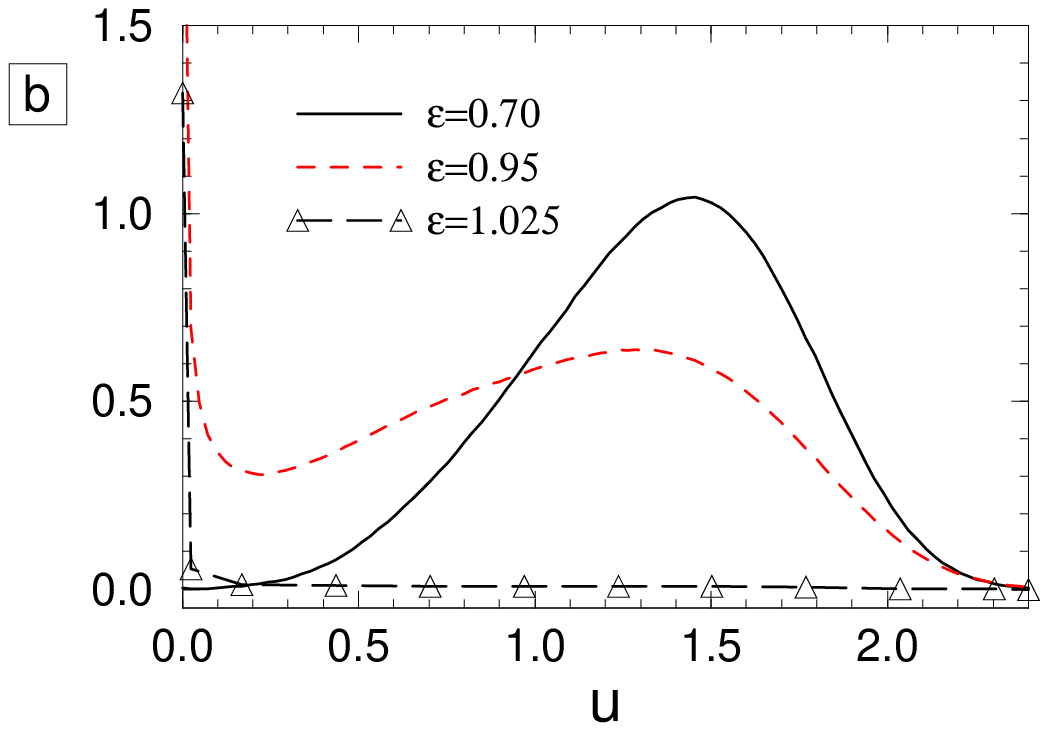,width=2.7in}
\caption[]{
(a) Fraction portion of absorbing phase $R$ 
as a function of $\epsilon$ and two different values of $a$. 
The data are the result of averaging over 40 realizations, 
for a time $t=900$. Other simulation parameters
as in \Fig{pdesim}.
(b) One-site probability distribution
$P(u)$ for different values of $\epsilon$. The ordinate axis of
the pdf corresponding to $\epsilon=1.025$ is scaled down $10$ times.
\label{numchar}}
\end{figure}

The transition from the active phase to the absorbing phase that
we have numerically characterized  above, can be described within
a Weiss--like mean field theory specially devised to study the
effect of fluctuations\cite{vand}. We first notice that, by a
suitable rescaling of the space variable $x$, \Equ{pde} can be
rewriting as: $u_t = f(u)  + D_0\, u_{xx} + g(u) \;\xi$ with 
$f(u) = -\frac{\partial \phi}{\partial u} $ and  $D_0\equiv
D/\epsilon^2$. We next consider a spatial discretization of the
field where $u_i=u(x_i,t)$.
One can write the multi--variate Fokker--Planck equation for the
set of variables $\{u_1,u_2 \dots \}$ which, after 
integration of all the $u_i$ variables except one, yields the 
equation for the stationary one-site pdf $P(u)$:
\begin{equation}
\partial_u\left[(-f(u) + 2 D_0\,(u-E(u)))P\right] + 
\partial^2_u (g^2(u) P)=0
\label{fokker}
\end{equation}
where $E(u)=\int  u'\; P(u'|u)\;du'$ is the steady-state
conditional average of the field at a nearest neighbor site of
a site in which the field takes the value $u$.
The above mentioned mean--field approximation 
takes $E(u) \simeq \;\bar{u}$, the yet unknown mean field value of $u$.
This is analogous to the traditional Weiss mean--field approach
in the theory of critical phenomena. We integrate
\Equ{fokker} to obtain the steady state pdf:
\begin{equation}
P(u;\bar{u}) = \frac{1}{Z\; g(u)^2}\, \exp
\int_0^u
\frac{f(v)-2D_0(v-\bar{u})}{g(v)^2}\; dv
\end{equation}
with $Z$ the normalization constant and where 
the dependence on the mean value
$\bar u$ has been explicited. The value of 
$\bar{u}$ arises from the consistency relation:
\begin{equation}
\bar{u}=\int_0^\infty v\,P(v;\bar{u})\,dv
\label{consistent}
\end{equation}
We can easily solve \Equ{consistent} in the no-coupling limit $D_0=0$,
where $P(u) \sim u^{-2(1+a)}$
which is non-normalizable (remember that $a>0$) so in fact we have
$P(u)=\delta(u)$, and hence $\bar{u}=0$, which describes the absorbing phase. 
On the other hand, 
the limit $D_0 =\infty$ can be treated using a saddle-point
expansion in $D_0$ which results in the equation $f(\bar{u})=0$
\cite{CommentStrato}. This equation coincides with the steady
state result of the deterministic analysis for a spatially
averaged field. Such analysis predicts a transition from the
absorbing state $u=0$ to the state $u=u_+$ at the Maxwell point
$a=a_M$. 
These limiting results show that the transition from STI to
absorbing appears in our parameter regime $0<a<a_M$ as a joint
effect of fluctuations {\em and} spatial coupling. We note that
other transition to an absorbing phase, recently studied in the
context of SPDE's, already exists for $D_0=0$\cite{muno98}.
\begin{figure}
\centerline{\epsfig{figure=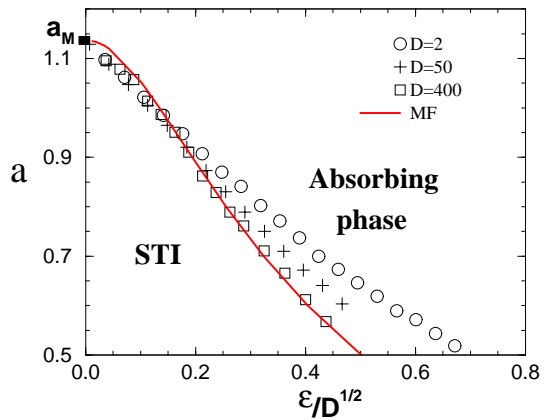,width=2.8in}}
\caption[]{Phase diagram of the noise induced transition from
STI to absorbing phase. The symbols correspond to numerical simulations
of \Equ{pde} for different values of $D$ and $\epsilon$. The solid
line comes from the solution of the mean-field consistency 
\Equ{consistent}\cite{CommentPhasedia}. 
\label{comparison}}
\end{figure}
In order to obtain results for intermediate values in the
coupling $D_0$ we solve numerically the consistency relation
\Equ{consistent}. In   \Fig{comparison} we show a phase diagram
where the mean field approach and the numerically determined
transition from \Equ{pde}, are compared.

A possible variation of our model is to include a finite
correlation length in the Gaussian white noise (\Fig{color}).
This modified model recreates the characteristic triangular
structures displayed by several two-component {\em
deterministic} reaction-diffusion
equations\cite{argen97,zimm97}.
Note that in this case the transition to the laminar state
occurs for non-zero velocity of the fronts, in analogy to what is
observed in deterministic models\cite{argen97}.  

\begin{figure}
\epsfig{figure=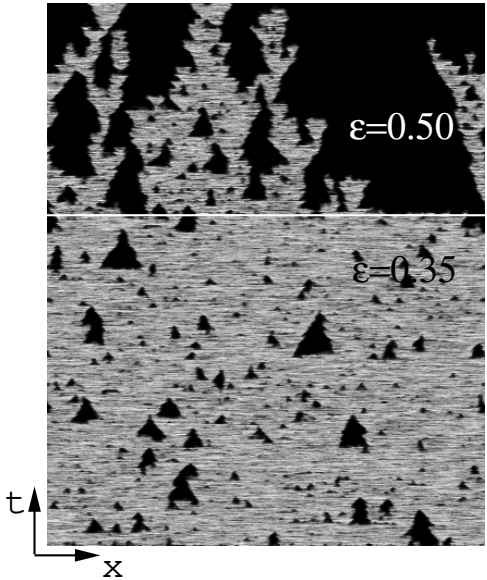,width=\sizeps}
\caption[]{Numeric integration of \Equ{pde} with periodic
boundary conditions and spatially correlated noise with 
$\langle\xi(x,t)\xi(x',t')\rangle \;=
\exp(-\frac{|x-x'|}{\gamma})\delta(t-t')$ for two noise intensities: 
$\epsilon=0.35$ for $t\in(0,150)$ and $\epsilon=0.50$ for
$t\in(150,250)$; 
$x\in(0,400)$ and $\gamma=8.0, D=2.0$.
\label{color}}
\end{figure}

As a conclusion, we have shown that STI can be modeled by means of
a simple SPDE. 
Our model  provides a new insight into
the complex behavior of the STI phase as a stochastic nucleation
of an absorbing metastable state. 
One of our main results is the occurrence of a noise induced transition where 
STI disappears in favor of an absorbing phase  at
sufficiently high noise intensity or low enough spatial coupling.
Notice that both, transitions to absorbing states
\cite{Tu97,muno98} and noise induced phase transition \cite{raul}
are known in the SPDE context.  The former phenomenon does
not require spatial coupling. The latter has been so far
associated with transitions between statistically homogeneous
stationary states. The noise induced transition shown here is a
genuine
consequence of stochasticity in extended systems.
It describes a transition between an absorbing phase and a
dynamically active and structured phase which is a form of STC.
Specific features of this active phase are described by
considering space-time configurations associated with the
individual realizations of the stochastic dynamics.
In summary we believe the SPDE approach to be general to explore
new generic mechanisms  of  spatiotemporal chaos.

We acknowledge helpful comments from J. M. Sancho and M. A.
Mu\~noz, and financial support from DGES (Spain) projects
PB94-0167, PB97-0141-C02-01. M.G.Z. is supported from a post-doctoral
grant of the MEC (Spain).

\end{twocolumns}


\begin{thebibliography}{99}
\bibitem{Itonumerics}
This SPDE is to be considered in the It\^o interpretation such that 
the noise term does not alter neither the location nor the
stability of the absorbing state at $u=0$.
\bibitem{CommentUniver}
This choice make our SPDE model in the multiplicative noise
universality class\cite{muno98}.
\bibitem{Rem_fronts}
Parameters are chosen such that nucleation events of the laminar
phase occur on a much longer time-scale than the one shown.
\bibitem{CommentStrato}
Note that Stratonovich interpretation has been used in
\cite{vand}, so that the $D_0\rightarrow \infty$ result amounts to
add the spurious drift term:  $f(u)+g(u) g'(u)/2=0$.  
Our results here do not follow from the effect of the
spurious drift.
\bibitem{CommentPhasedia} 
For each value of $D_0$ the consistency equation
\Equ{consistent} for $0<a<a_M$ gives 3 solutions which
continuously match the solutions given by $f(\bar u)=0$ as 
$D_0\rightarrow \infty$.
The relative stability can be analyzed by means of a Maxwell--like
construction, which predicts a transition at the solid line in
\Fig{comparison}.
\end{thebibliography}
\end{document}